\documentstyle[12pt,epsf]{article}
\begin{document}
\begin{center}
{\bf Effect of Nonmagnetic Impurities on the Magnetic Resonance
Peak in $\bf YBa_2 Cu_3 O_7$}

\vspace{.5in}

\noindent  H.F. Fong$^{(1)}$, P. Bourges$^{(2)}$, Y. Sidis$^{(2)}$,
L.P. Regnault$^{(3)}$, J. Bossy$^{(4)}$,\\ A. Ivanov$^{(5)}$,
D.L. Milius$^{(6)}$, I.A. Aksay$^{(6)}$, and B. Keimer$^{(1,7)}$

\end{center}

\vspace{.5in}

\begin{tabular}{lp{5in}}
1& Department of Physics, Princeton University, Princeton, NJ 08544
USA\\
2& Laboratoire L\'eon Brillouin, CEA-CNRS, CE Saclay, 91191 Gif sur 
Yvette, France\\
3& CEA Grenoble, D\'epartement de Recherche Fondamentale sur 
la mati\`ere Condens\'ee, 38054 Grenoble cedex 9, France\\
4& CNRS-CRTBT, BP 156, 38042 Grenoble Cedex 9, France\\
5 & Institut Laue-Langevin, 156X, 38042 Grenoble Cedex 9, France\\
6 & Department of Chemical Engineering, Princeton University, 
Princeton, NJ 08544 USA\\
7& Max-Planck-Institut f\"ur Festk\"orperforschung, D-70569 Stuttgart, Germany
\end{tabular}

\vspace{.8in}

\begin{center}
ABSTRACT
\end{center}

\vspace{.1in}

\noindent The magnetic excitation spectrum of a $\rm YBa_2 Cu_3 O_7$ 
crystal containing 0.5\% of nonmagnetic (Zn) impurities has been determined 
by inelastic neutron scattering. Whereas in the pure system a sharp resonance 
peak at $E \simeq 40$ meV is observed exclusively below the 
superconducting transition temperature $\rm T_c$, 
the magnetic response in the Zn-substituted system
is broadened significantly and vanishes at a temperature much {\it higher} than $\rm T_c$. 
The energy-integrated spectral weight observed near ${\bf q}= (\pi,\pi)$ increases
with Zn substitution, and only about half of the spectral weight is removed at $\rm T_c$.

\clearpage

The magnetic resonance peak is a sharp collective excitation at
an energy of 40 meV and wavevector ${\bf q}_0 = (\pi,\pi)$ in the
superconducting state of $\rm YBa_2 Cu_3 O_7$ 
\cite{rossat91}-\cite{regnault98}. The peak is also
observed at lower energies in underdoped $\rm YBa_2 Cu_3 O_{6+x}$
\cite{dai96,fong97,bourges97}.
As the existence of this peak requires d-wave superconductivity,
it demonstrates that magnetic neutron scattering is a phase sensitive
probe of superconductivity \cite{fong95}. The peak also provides
important clues to the microscopic mechanism of high temperature
superconductivity: It does not appear in the Lindhard susceptibility
of a noninteracting band metal \cite{mazin95}, and the interactions
responsible for the enhancement of the band susceptibility are presumably
the same as the ones that drive superconductivity.
Several enhancement mechanisms have thus been suggested: band structure
singularities \cite{band}, antiferromagnetic interactions 
\cite{antiferro}, and interlayer tunneling \cite{chakravarty97}. 
Other models of the resonance peak 
\cite{zhang95,pines96,assaad98} appeal directly to
the parent antiferromagnetic insulator $\rm YBa_2 Cu_3 O_6$, where
low energy spin waves of spectral weight comparable to the resonance
peak are observed near ${\bf q}_0 = (\pi,\pi)$.

Clearly, more experimental data are needed to differentiate between
these theories. Here we present neutron scattering measurements
of the spin dynamics of a $\rm YBa_2 Cu_3 O_7$ single crystals in
which a small number of nonmagnetic zinc ions replace copper ions. Zn substitution 
introduces minimal structural disorder, substitutes
for copper in the $\rm CuO_2$ planes \cite{maeda89}, and does not modify
the hole concentration substantially \cite{alloul91}. The Zn ions are known to induce local
magnetic moments on neighboring Cu sites \cite{mahajan} which are associated with 
low energy magnetic excitations \cite{sidis96,kakurai93,matsuda93}. It has further been 
shown that Zn impurities scatter conduction electrons near the unitary limit 
and rapidly suppress the superconducting transition temperature, $\rm T_c$ \cite{transport}. 
The new data reported
here demonstrate a broadening of the spin excitation spectrum in the 
presence of a minute amount of Zn impurities. Instead of
disappearing in the normal state as in zinc-free $\rm YBa_2 Cu_3 O_7$, 
the broadened intensity now persists well above $\rm T_c$. These results are 
surprising and were not anticipated by any of the theoretical models of the resonance peak.

A single crystal of composition $\rm YBa_2 (Cu_{0.995} Zn_{0.005})_3 O_7$ and volume
$\rm 1.7 cm^3$ was prepared by a
method described previously \cite{fong96}. The crystal was annealed
under oxygen flow at $\rm 600^\circ C$ for 14 days, a procedure that resulted in
a $\rm T_c = 93K$ in Zn-free crystals synthesized by the same 
method \cite{fong96}. After the heat 
treatment, the crystal showed $\rm T_c = 87 K$ width a width of about 5K,
consistent with earlier reports on Zn-substituted, fully
oxygenated $\rm YBa_2 Cu_3 O_7$ \cite{transport}. 

The measurements were taken at the IN8 triple axis spectrometer
at the Institut Laue Langevin, Grenoble, France. Preliminary data were
also taken at the BT2 spectrometer at the NIST research reactor.
The IN8 beam optics included
a vertically focusing Cu (111) monochromator, and a horizontally focusing 
pyrolytic graphite (002) analyser which selected a fixed final
energy of 35 meV. A pyrolytic graphite filter was inserted
into the scattered beam in order to eliminate higher-order contamination.
The sample was attached to the cold finger of a closed cycle helium
refrigerator mounted on a two-circle goniometer. Data were taken with 
the crystal in two different orientations where
wave vectors of the forms ${\bf Q} = (H,H,L)$ and $(3H,H,L)$ were accessible. 
[Throughout this article, the wave vector ${\bf Q} = (H, K, L)$
is indexed in units of the reciprocal lattice vectors $2 \pi/a 
\sim 2\pi/b  \sim 1.63 {\rm \AA}^{-1}$ and $2\pi/c \sim 0.53 {\rm
\AA}^{-1}$. In this notation, the $(\pi,\pi)$ point corresponds
to $(\frac{h}{2},\frac{k}{2})$ with $h$ and $k$ integers.]

As described in detail elsewhere \cite{bourges96,fong96}, the imaginary part 
of the dynamical magnetic susceptibility, $\chi''({\bf Q},\omega)$,
can be separated from phonon scattering with the aid of 
lattice vibrational calculations, and by studying the momentum, temperature and doping 
dependence of the neutron scattering cross section. In pure (Zn-free) 
$\rm YBa_2 Cu_3 O_{6+x}$, this procedure was verified
by measurements with polarized neutron beams for some scattering configurations
in the energy range
covered by the present study, 10 meV through 50 meV \cite{fong96,fong97}. 
Since the changes in the phonon spectrum induced by 0.5\% Zn-substitution are insignificant,
the data analysis procedures developed for pure $\rm YBa_2 Cu_3 O_{6+x}$ carry 
over directly to the sample investigated here. 

Fig. 1 shows representative constant-energy scans taken in the $(H, H, L)$ zone. As in pure
$\rm YBa_2 Cu_3 O_7$, the magnetic scattering is confined to a window around 40 meV and 
${\bf q}_0=(\pi,\pi)$, with no magnetic scattering observed above background at low energies. The
magnetic scattering around 40 meV also exhibits the sinusoidal modulation in $L$ discussed in 
detail previously \cite{rossat91}-\cite{fong96},\cite{reznik96}. An upper limit
of 1/3 of the maximum intensity can be placed on the intensity at the minimum of the
modulation, as in the pure system. These and similar scans
taken in the $(3H, H, L)$ zone were put on an absolute
unit scale by normalizing to the phonon spectrum following Ref. \cite{fong96}, using 
the same definition of the spin susceptibility as Ref. \cite{ybco6.5} (Fig. 2a). 
The constant-energy scans were fitted to Gaussian profiles whose amplitude is
plotted in Fig. 2b as a function of energy. 

The overall shape of the magnetic spectrum 
of $\rm YBa_2 (Cu_{0.995} Zn_{0.005})_3 O_7$, being concentrated around a characteristic 
energy 40 meV and wave vector ${\bf q}_0=(\pi,\pi)$, is
clearly reminiscent of the resonance peak in the pure system. 
The spectral weight increases at low temperatures, as does the 
resonance peak in pure $\rm YBa_2 Cu_3 O_7$. However, there are also 
substantial differences between the pure and Zn-substituted systems. First, while the magnetic 
resonance peak in the pure system is very sharp in energy, the magnetic 
response in the Zn-substituted sample is substantially broadened. 
This is apparent in the temperature difference spectrum
(Fig. 2c), which gives a more detailed picture of the energy range near 40 meV. (In order to 
obtain better counting statistics, the high temperature cross section at 
constant wave vector was subtracted from the low temperature cross section 
at the same wave vector, without a full {\bf Q}-scan at each 
energy. Since the phonon scattering is temperature independent in this 
energy and temperature range, only magnetic scattering contributes to 
the difference spectrum.) The full width at half maximum of the difference spectrum
is $\Delta E \sim 10$ meV, much broader than the instrumental resolution ($\sim 5$ meV), 
yielding an intrinsic energy width of $\sim 8.5$ meV. The width of the unsubtracted spectrum
in Fig. 2b is also consistent with this value.
Within the errors, the full width at half maximum in 
{\bf Q}-space, $\Delta Q \sim 0.25 {\rm \AA}^{-1}$ at $E=39$ meV, is identical to 
the resonance width in the pure system (Fig. 1).

A series of constant-energy scans at energies 39 meV and 35 meV were carried out
at temperatures up to $\sim 300$K and fitted to Gaussian profiles. The fitted amplitudes for 39 meV
are plotted in Fig. 3; the data for 35 meV track those of Fig. 3 to within the errors. 
This figure fully reveals an even more
dramatic difference of the spectra in the pure and Zn-substituted systems, already indicated in 
Figs. 2a and b. Whereas the magnetic response in the pure system (restricted to a single resonance peak) 
disappears in the normal state \cite{fong95,bourges96,fong96}, Fig. 3 shows that in
$\rm YBa_2 (Cu_{0.995} Zn_{0.005})_3 O_7$ the magnetic spectral weight 
actually persists up to $\sim 250$K. Furthermore, in both underdoped and optimally doped systems, 
the magnetic resonance peak follows a sharp, order parameter-like curve below $\rm T_c$ 
(Refs. \cite{bourges96}-\cite{bourges97}). By contrast, there is at most a weak inflection
point near $\rm T_c$ in the Zn-substituted system. The influence of superconductivity on the
spin excitations, which is so clearly apparent in pure $\rm YBa_2 Cu_3 O_{6+x}$, is thus almost
completely obliterated by 0.5\% Zn substitution.

A further important comparison between the pure and Zn-substituted materials is made possible
by the absolute unit calibration. Since the resonance peak in the pure system 
is very sharp and comparable to the instrumental energy resolution, the appropriate 
quantity to compare is the energy-integrated magnetic spectral weight,
$\int d \omega \; \chi'' ({\bf q}_0,\omega)$, in the energy range probed by the neutron experiment.
This quantity is 2.2 $\pm$ 0.5 $\mu_B^2$ at low temperatures in 
$\rm YBa_2 (Cu_{0.995} Zn_{0.005})_3 O_7$, as compared to 1.6 $\pm$ 0.5 $\mu_B^2$ in 
$\rm YBa_2 Cu_3 O_7$ \cite{footnote}. (Note, however, that in 
$\rm YBa_2 (Cu_{0.995} Zn_{0.005})_3 O_7$ only
half of this intensity is removed upon heating to $\rm T_c$, whereas in the pure
system no magnetic intensity is observable above $\rm T_c$.)
As nonmagnetic impurities are added, the total energy-integrated spectral weight 
around $(\pi,\pi)$ therefore {\it increases} in the energy range probed by the neutron experiment, 
implying that zinc restores antiferromagnetic 
correlations. In this respect, the Zn-substituted system resembles the 
underdoped pure system (x $< 0.95$) where a normal state 
antiferromagnetic contribution exists \cite{rossat91,regnault,regnault98}. Surprisingly, in the Zn-doped system this additional intensity appears in the same energy and wave vector range as the resonance peak 
in the pure system. 

It is also interesting to compare the present data to previous neutron scattering work on more 
heavily Zn substituted cuprate superconductors \cite{sidis96}-\cite{matsuda93}. 
The enhanced low energy excitations near ${\bf q}_0=(\pi,\pi)$ reported for these 
materials were not observed in our very lightly Zn-substituted sample (bottom panel in Fig. 1). 
However, at higher energies a 2\% Zn-substituted, 
fully oxygenated $\rm YBa_2 Cu_3 O_7$ sample investigated by Sidis {\it et al.} \cite{sidis96} 
exhibits a spectral distribution closely similar to 
the one shown in Fig. 2. The temperature evolution of the magnetic intensity \cite{regnault98} 
is also consistent with the one reported here (Fig. 3). 

In summary, the effect of substituting one out of 200 copper atoms by nonmagnetic impurities
is dramatic. The total spectral weight
near ${\bf q}_0=(\pi,\pi)$ actually increases and persists to higher temperatures while remaining
centered around 40 meV. On the other hand, the characteristic features of the resonance peak 
({\it i.e.}, its sharpness in energy and its coupling to superconductivity) are obliterated. It is
worth noting that in the underdoped regime, where the normal-state susceptibility is also
enhanced with respect to $\rm YBa_2 Cu_3 O_7$, the resonance peak remains sharp and coupled to 
superconductivity \cite{dai96}-\cite{bourges97}. This aspect thus seems to be a 
manifestation of a delicate
coherence that is very easily disrupted by disorder. While none of the theories of 
the resonance peak \cite{mazin95}-\cite{pines96} has anticipated this
behavior, it is reminiscent of the extreme susceptibility of collective-singlet ground
states in quasi-one dimensional systems (realized, for instance, in spin-Peierls and spin ladder
materials) to nonmagnetic impurities. A microscopic analogy 
between both systems was pointed out by Fukuyama and coworkers \cite{fukuyama96}, but
its consequences for the spin excitations have not yet been evaluated. Viewed from a different
angle, a gradual buildup of spectral weight below $\rm T \sim 250$K (as shown in Fig. 3) is also
observed in underdoped $\rm YBa_2 Cu_3 O_{6+x}$, where it is centered around a somewhat lower
energy (20-30 meV) and goes along with the opening of the ``spin pseudo-gap'' 
\cite{regnault,ybco6.5}. The strong temperature evolution in the normal states of
both underdoped and disordered $\rm YBa_2 Cu_3 O_{6+x}$ is obviously closely 
related to the resonance peak and should be part of a comprehensive 
theoretical description of the spin dynamics of the cuprates.

\vspace{.2in}

\noindent {\bf Acknowledgments}\\
We are grateful for technical assistance provided by D. Puschner.  
The work at Princeton University was supported by the National 
Science Foundation under Grant No. DMR-9400362, and by the Packard 
and Sloan Foundations.

\clearpage

\subsection*{Figure Captions}
\begin{enumerate}

\item Constant-energy scans at 39 meV and 10 meV through ${\bf Q}=(\frac{1}{2},\frac{1}{2},L)$ for 
$\rm YBa_2 (Cu_{0.995} Zn_{0.005})_3 O_7$. The line in the upper panel is the results of a 
Gaussian fit. The bar gives the instrumental {\bf Q}-resolution. Because of the good
energy resolution ($\sim 5$ meV), the 42.5 meV phonon \cite{fong95} makes only a weak
contribution ($\leq 10$\%) to the upper scan.

\item a) Constant-energy scans through (1.5, 0.5, -1.7), background corrected and converted
to absolute units. The bar gives the instrumental {\bf Q}-resolution.
b) Peak dynamical susceptibility at ${\bf q}_0=(\pi,\pi)$ extracted 
from fits to constant-energy profiles (panel a and Fig. 1). The line is a Gaussian with 
the same width as the difference spectrum in panel c.
c) More detailed spectrum around 40 meV. The data around 100K ($\rm > T_c$)
were subtracted from the low temperature data. The bar gives the instrumental
energy resolution, and the line is the result of a fit to a Gaussian. All data are given in absolute units. 
A $\sim 30$\% overall systematic error in
the absolute unit calibration is not included in the error bars.

\item Temperature dependence of the dynamical susceptibility at ${\bf q}_0=(\pi,\pi)$ and 
at the peak energy of the spectrum ($\sim$ 39 meV), in absolute units. The closed circles are the 
fitted amplitudes of constant-energy 
scans, the open circles are the peak count rates.
\end{enumerate}


\clearpage

\begin{thebibliography}{99}

\bibitem{rossat91} J. Rossat-Mignod {\it et al.}, Physica C {\bf 185-189}, 86 (1991).
 
\bibitem{mook93} H.A. Mook {\it et al.}, Phys. Rev. Lett. {\bf 70}, 3490 (1993).

\bibitem{regnault} L.P. Regnault {\it et al.}, Physica C, {\bf 235-240}, 59, (1994); 
Physica B, {\bf 213\&214}, 48, (1995). 

\bibitem{fong95} H.F. Fong {\it et al.}, Phys. Rev. Lett. {\bf 75}, 316 (1995).

\bibitem{bourges96} P. Bourges, L.P. Regnault, Y. Sidis and C. Vettier, 
Phys. Rev. B {\bf 53}, 876 (1996).

\bibitem{fong96} H.F. Fong {\it et al.}, Phys. Rev. B. {\bf 54}, 6708 (1996).

\bibitem{regnault98} L.P. Regnault {\it et al.} in {\it Neutron Scattering 
in Layered Copper-Oxide Superconductors }, 
Edited by A. Furrer, (Kluwer, Amsterdam, 1998), p. 85.

\bibitem{dai96} P. Dai {\it et al.}, Phys. Rev. Lett {\bf 77}, 5425 (1996).

\bibitem{fong97} H.F. Fong, B. Keimer, D.L. Milius and I.A. Aksay,
Phys. Rev. Lett. {\bf 78}, 713 (1997).

\bibitem{bourges97} P. Bourges {\it et al.}, Europhys. Lett. {\bf 38}, 313 (1997).

\bibitem{mazin95} I.I. Mazin and V.M. Yakovenko, Phys. Rev. Lett.
{\bf  75}, 4134 (1995).

\bibitem{band} N. Bulut and D.J. Scalapino, Phys. Rev. B 
{\bf 53}, 5149 (1996); G. Blumberg, B.P. Stojkovic and 
M.V. Klein, {\it ibid.} {\bf 52}, 15741 (1995); A.A.
Abrikosov, {\it ibid.} {\bf 57}, 8656 (1998).

\bibitem{antiferro} D.Z. Liu, Y. Zha and K. Levin, Phys. Rev. Lett.
{\bf 75}, 4130 (1995); F. Onufrieva, Physica C {\bf 251}, 348 (1995);
A.J. Millis and H. Monien, Phys. Rev. B {\bf 54}, 16172 (1996). 
	
\bibitem{chakravarty97} L. Yin, S. Chakravarty and P.W. Anderson, 
Phys. Rev. Lett. {\bf 78}, 3559 (1997).

\bibitem{zhang95} E. Demler and S.C. Zhang, Phys. Rev. Lett. 
{\bf 75}, 4126 (1995); S.C. Zhang, Science {\bf 275}, 1089 (1997).

\bibitem{pines96} Y. Zha, V. Barzykin and D. Pines, Phys. Rev. B 
{\bf 54}, 7561 (1996); D.K. Morr and D. Pines, Report No.
cond-mat/9805107.

\bibitem{assaad98} F.F. Assaad and M. Imada, Report No. cond-
mat/9711172.

\bibitem{maeda89} G. Xiao {\it et al.}, Nature {\bf 332}, 238 (1988);
H. Maeda {\it et al.}, Physica C {\bf 157}, 483 (1989).

\bibitem{alloul91} H. Alloul {\it et al.}, Phys. Rev. Lett. {\bf 67}, 3140 (1991).

\bibitem{mahajan} V. A. Mahajan {\it et al.}, Phys. 
Rev. Lett. {\bf 72}, 3100 (1994).

\bibitem{sidis96} Y. Sidis {\it et al.}, Phys. Rev. B {\bf 53}, 6811 (1996).

\bibitem{kakurai93} K. Kakurai {\it et al.}, Phys. Rev. B {\bf 48}, 3485
(1993).

\bibitem{matsuda93} M. Matsuda {\it et al.}, J. Phys. Soc. Jpn. {\bf 62}, 443
(1993).

\bibitem{transport} T.R. Chien, Z.Z. Wang and N.P. Ong, Phys. Rev. Lett. {\bf 67}, 2088 (1991);
D.A. Bonn {\it et al.}, Phys. Rev. B {\bf 50}, 4051 (1994); 
Fukuzumi {\it et al.}, Phys. Rev. Lett.{\bf 76}, 684 (1996).

\bibitem{reznik96} D. Reznik {\it et al.},
Phys. Rev. B {\bf 53}, R14741 (1996); S.M. Hayden {\it et al.}, {\it ibid.}
{\bf 54}, R6905 (1996).

\bibitem{ybco6.5}  P. Bourges {\it et al.}, Phys. Rev. B  {\bf 56},
R11439 (1997).

\bibitem{footnote} Note that if the susceptibility is energy integrated in the energy
range probed by the experiment {\it and} averaged over the two-dimensional Brillouin zone, 
the resulting numbers, $\int d \omega \int d^2 q \; \chi'' ({\bf q}_0,\omega) \; / \; 
\int d^2 q = 0.058 \mu_B^2$ in $\rm YBa_2 (Cu_{0.995} Zn_{0.005})_3 O_7$
and $0.043 \mu_B^2$ in $\rm YBa_2 Cu_3 O_7$ come out much smaller than
the corresponding number $\frac{\pi}{3} s(s+1) g^2 \mu_B^2$ required by the
total moment sum rule for an insulating $s=1/2$ antiferromagnet. In deriving these numbers, 
we used the full width at half maximum in momentum space, which is 0.25 ${\rm \AA}^{-1}$ in both materials. 
Note also that in underdoped materials where the spin excitation spectrum is much broader in energy, 
it is often convenient to quote the Brillouin zone averaged (local) susceptibility 
$\int d^2 q \; \chi'' ({\bf q}_0,\omega) \; / \; \int d^2 q$ {\it without} integrating over energy \cite{ybco6.5}. 

\bibitem{fukuyama96} N. Nagaosa {\it et al.}, J. Phys. Soc. Jpn. {\bf 65}, 3724 (1996); H. Fukuyama, T. 
Tanimoto, and M. Saito, {\it ibid.} {\bf 65}, 1182 (1996).

\end{thebibliography}
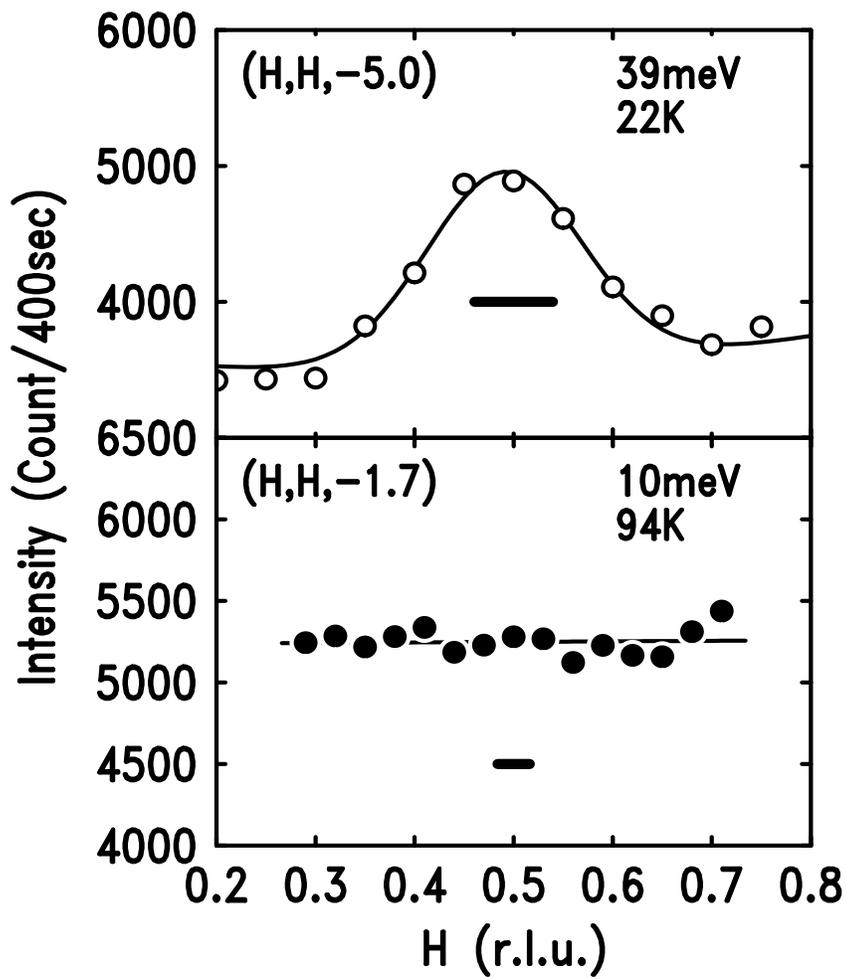
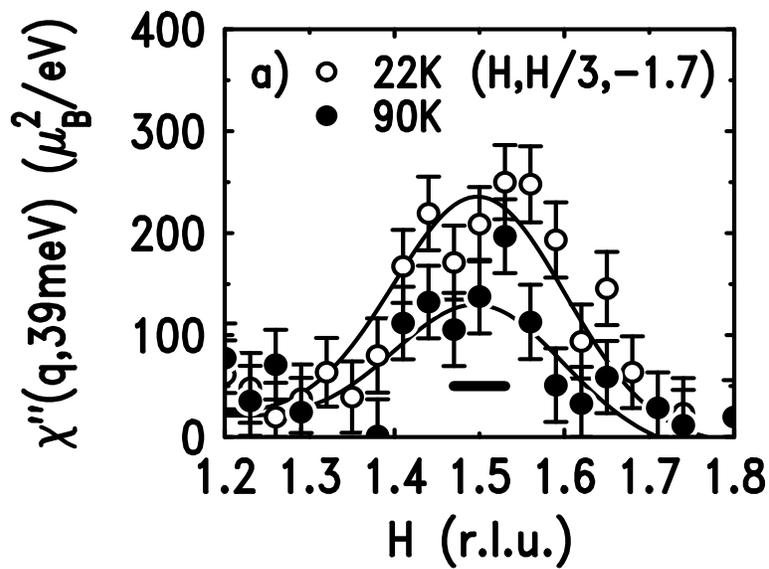
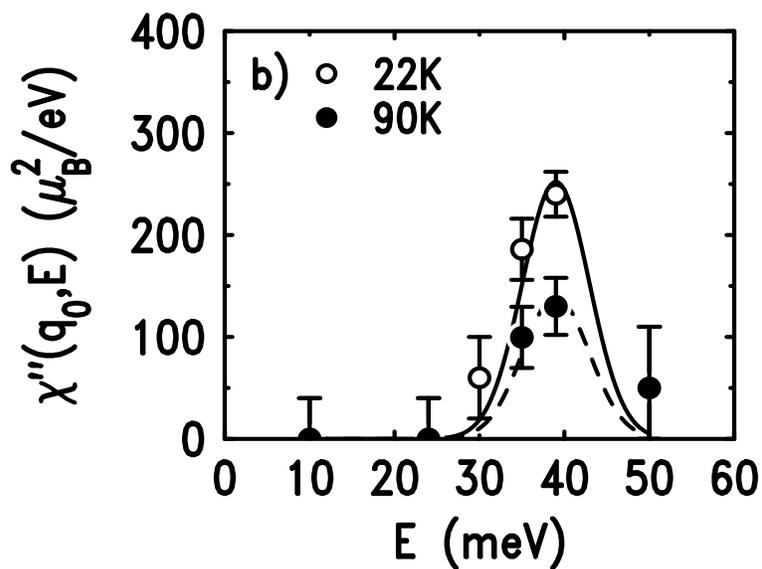
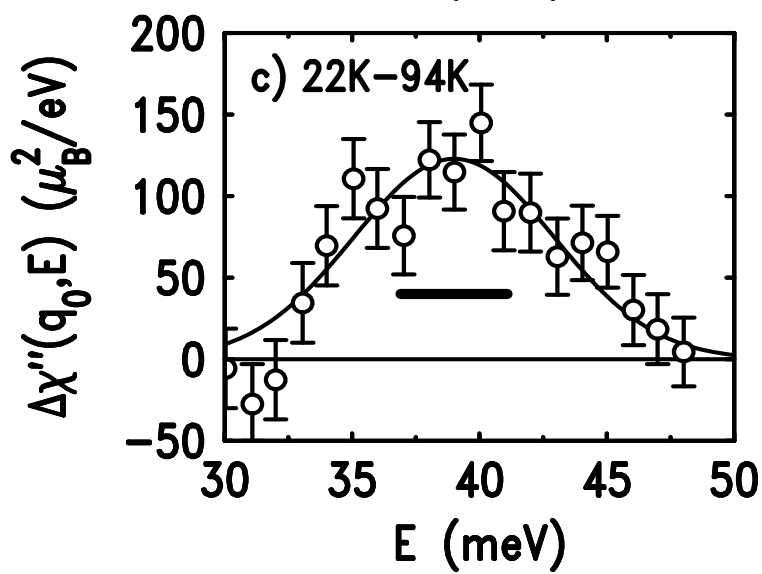
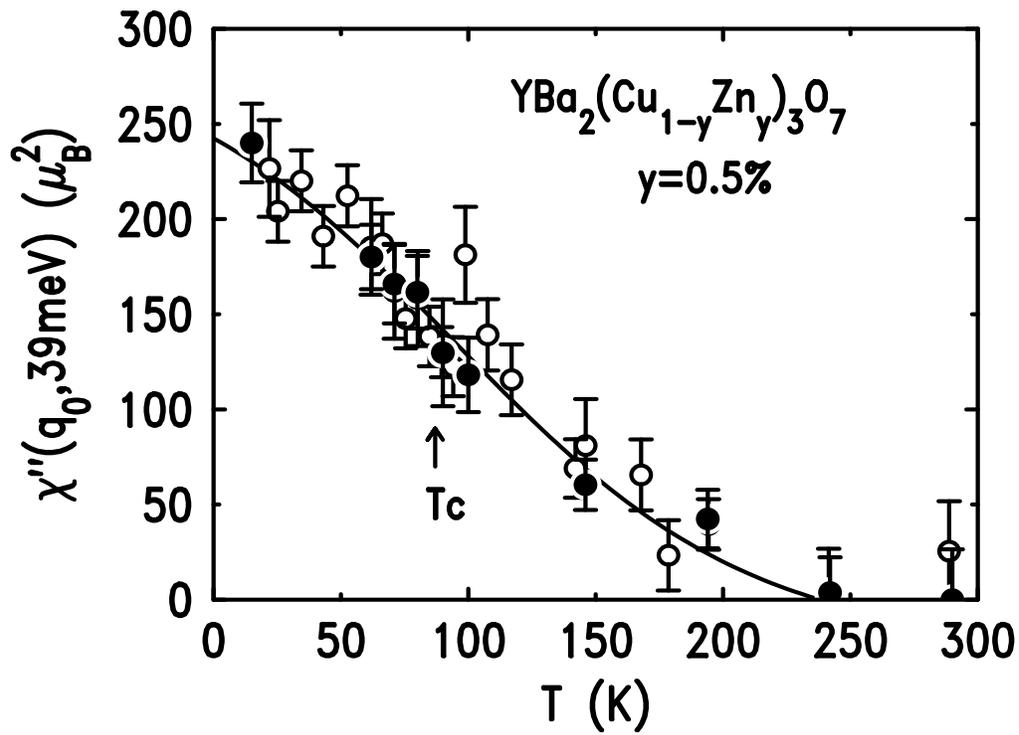
\end{document}